\documentclass{aa} 

\usepackage{graphicx}
\usepackage{makecell}
\usepackage{hyperref}
\hypersetup{linkcolor=blue,citecolor=blue,colorlinks=true,pdfborderstyle={}}
\usepackage{natbib}
\usepackage{xcolor}
\usepackage{xspace}
\usepackage{xcolor}
\usepackage{amsmath}
\usepackage{amssymb}
\usepackage{multirow}
\usepackage{graphicx}
\usepackage{booktabs}
\usepackage{adjustbox}
\usepackage{subcaption}
\usepackage{enumerate}
\usepackage{enumitem}
\usepackage{float}

\bibpunct{(}{)}{;}{a}{}{,} 
\usepackage{txfonts}
\begin{document}

    \title{Possible contributions of two nearby blazars  to the NGC~4151 neutrino hotspot}

   \author{A. Omeliukh
          \inst{1}\fnmsep\thanks{\email{omeliukh@astro.rub.de}}
          \and 
          S. Barnier \inst{2}
          \and
          Y. Inoue\inst{2, 3, 4}
          }

   \institute{
   Ruhr University Bochum, Faculty of Physics and Astronomy, Astronomical Institute (AIRUB),  Universitätsstraße 150, 44801 Bochum, Germany
         \and
    Osaka University, Department of Earth and Space Science, Graduate School of Science, 1-1 Machikaneyama, Toyonaka, Osaka 560-0043,
Japan
    \and
    Interdisciplinary Theoretical \& Mathematical Science Program (iTHEMS), RIKEN, 2-1 Hirosawa, Saitama 351-0198, Japan
    \and
    Kavli Institute for the Physics and Mathematics of the Universe (WPI), UTIAS, The University of Tokyo, Kashiwa, Chiba 277-8583, Japan
    }

   \date{Received XX; accepted XX}

 
  \abstract
   {The origin of the high-energy astrophysical neutrinos discovered by IceCube remains unclear, with both blazars and Seyfert galaxies emerging as potential sources. Recently, the IceCube Collaboration reported a ${\sim}{3}\sigma$ neutrino signal from the direction of a nearby Seyfert galaxy NGC 4151. However, two gamma-ray loud BL Lac objects, 4FGL~1210.3+3928 and 4FGL~J1211.6+3901, lie close to NGC~4151, at angular distances of 0.08$^\circ$ and 0.43$^\circ$, respectively.}
   {We investigate the potential contribution of these two blazars to the observed neutrino signal from the direction of NGC 4151 and assess their detectability with future neutrino observatories.}
   {We model the multi-wavelength spectral energy distributions of both blazars using a self-consistent numerical radiation code, AM$^3$. We calculate their neutrino spectra and compare them to the measured NGC 4151 neutrino spectrum and future neutrino detector sensitivities.}
   {The SED of 4FGL~1210.3+3928 revealed a feature that cannot be explained with a purely leptonic model suggesting the presence of protons in the jet. Our model predicts neutrino emission peaking above $\sim$10$^{17}$ eV with fluxes of ${\sim}10^{-12}~\mathrm{erg~cm^{-2}~s^{-1}}$ for this source. The SED of 4FGL~J1211.6+3901 can be explained with both leptonic and leptohadronic models. The contribution of these two blazars to the $\sim$10 TeV neutrino signal observed from the direction of NGC~4151 can only be minor. Still, future radio-based neutrino telescopes such as IceCube-Gen2's radio array and GRAND may be able to detect high-energy neutrinos from these two potential neutrino sources.
    }
   {}


   \maketitle
%

\section{Introduction}

More than ten years ago, the IceCube Neutrino Observatory discovered a diffuse flux of high-energy astrophysical neutrinos (\citealt{doi:10.1126/science.1242856}), the nature of which still remains elusive. One of the most plausible candidates are blazars, a rare and energetic subclass of active galactic nuclei (AGNs). AGNs are powered by the accretion of matter onto supermassive black holes at the centers of galaxies. Blazars, distinguished by their relativistic jets aligned close to our line of sight, are natural candidates for neutrino emitters due to their role as powerful cosmic-ray accelerators. The blazar TXS 0506+056 was the first source associated with high-energy neutrino emission (\citealt{doi:10.1126/science.aat1378}). IceCube has subsequently detected high-energy neutrino events in spatial and temporal coincidence with increased activity of several other individual blazars, among which are PKS~1424-41 (\citealt{2016NatPh..12..807K}; \citealt{2017ApJ...843..109G}), GB6~J1040+0617 (\citealt{2019ApJ...880..103G}), PKS~1502+106 (\citealt{Franckowiak_2020,Rodrigues_2021}), and PKS~0735+178 (\citealt{10.1093/mnras/stac3607}).

Recently, the landscape of potential neutrino sources has expanded beyond blazars to include other classes of AGNs. The IceCube Collaboration reported a 4.2$\sigma$ signal from the nearby Seyfert galaxy, NGC~1068 \citep{2020PhRvL.124e1103A, 2022Sci...378..538I}. Seyfert galaxies, characterized by weak or absent jet activity, represent a distinct class of AGNs. Their coronal activity could produce gamma-ray deficit neutrinos \citep[see e.g.,][]{Stecker:1991vm, Kalashev:2014vya, Inoue:2019fil, Inoue:2019yfs, Murase:2019vdl, Kheirandish:2021wkm, Gutierrez:2021vnk, Eichmann:2022lxh, Ajello_2023, Mbarek:2023yeq, Fiorillo:2023dts, 2024NatAs...8.1077P, Murase_2024, Das_2024}. This potential has been further supported by additional observational studies (\citealt{2024arXiv240607601A,PhysRevLett.132.101002, sommani2024100}). The IceCube collaboration has recently announced a ${\sim}3\sigma$ evidence for $\sim$10 TeV neutrino emission from the direction of another Seyfert galaxy, NGC~4151 \citep{2024arXiv240607601A, 2024arXiv240606684A}. This nearby Seyfert galaxy, located at a distance of $d = 15.8$~Mpc \citep{Yuan_2020}, provides an intriguing case study for investigating neutrino emission from Seyfert galaxies. Both corona and jet contributions of NGC~4151 itself to the neutrino signals have been studied in the literature (\citealt{Inoue:2023bmy, Murase:2023ccp}, but see also \citealt{2023arXiv230303298P}).

\begin{figure}
   \centering
   \includegraphics[width=8cm]{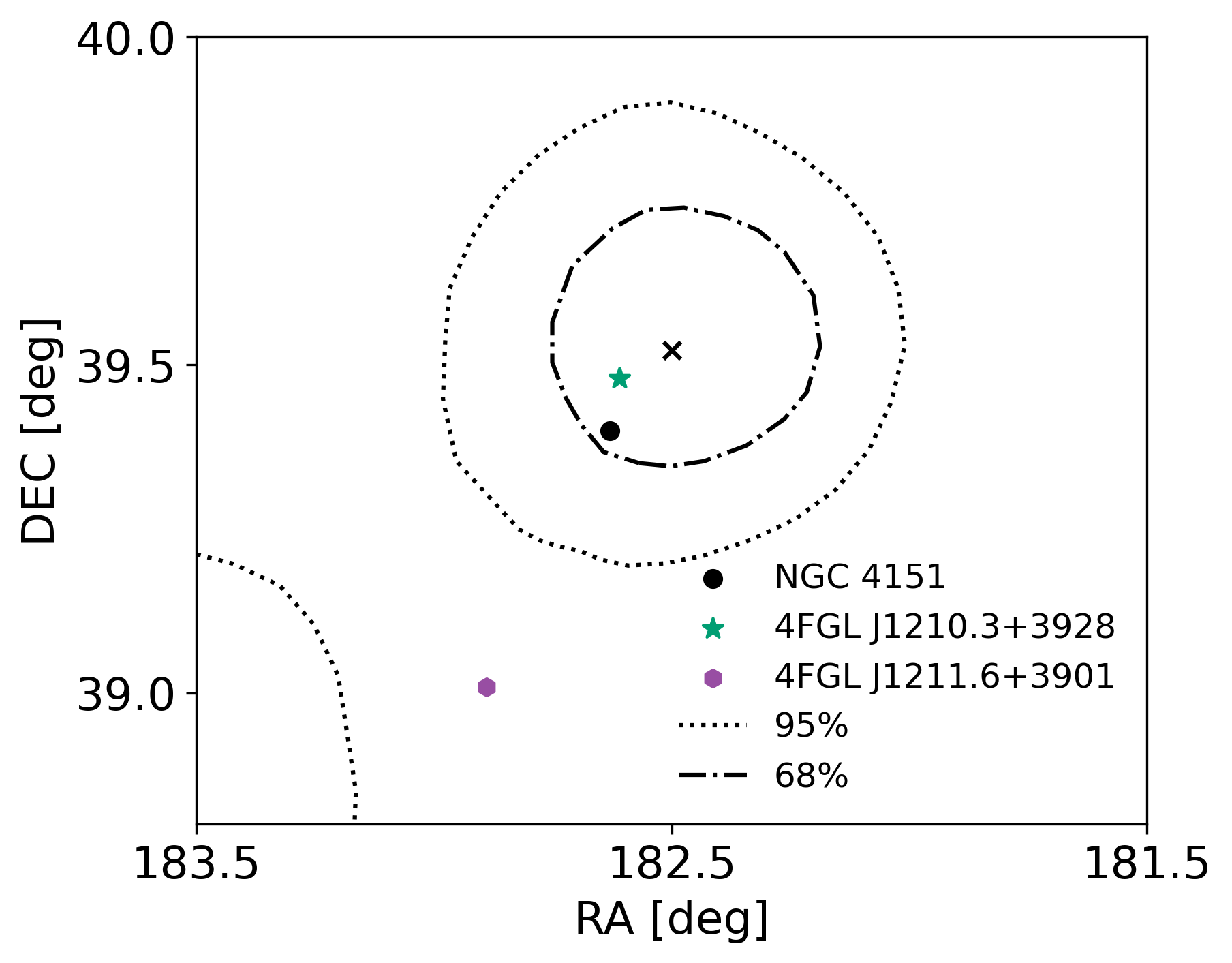}
      \caption{The location of 4FGL~J1210.3+3928 and 4FGL~J1211.6+3901 with respect to NGC~4151. The black cross, dash-dotted, and dotted black lines correspond respectively to the best-fit location of the neutrino source and its 68\% and 95\% confidence regions \citep{2024arXiv240606684A}.} 
         \label{fig:skymap}
   \end{figure}

However, two gamma-ray loud blazars, 4FGL~J1210.3+3928 and 4FGL~J1211.6+3901, are located 0.08$^\circ$ and 0.43$^\circ$ respectively from NGC~4151. 4FGL~J1210.3+3928 lies within the 68\% confidence region of the neutrino excess attributed to NGC~4151 \citep{2024arXiv240606684A} as shown in Fig.~\ref{fig:skymap}, making it a natural candidate for a contributing source. This source has also been associated with a neutrino hotspot in \cite{Buson:2023irp}. While 4FGL~J1211.6+3901 is located outside the 95\% confidence contour, we note that these contours depend on the assumed neutrino spectrum and can vary with  different spectral indices (including both  statistical and systematic uncertainties, \citealp[see Fig.~8 in][]{2024arXiv240606684A}). Given this potential variation in the contour shape and the fact that 4FGL~J1210.3+3928 and 4FGL~J1211.6+3901 are the only two nearby gamma-ray loud sources \citep{Murase:2023ccp}, we include both objects in our analysis.  

Since blazars have been promising neutrino emitters, there might be a potential contribution from these two blazars to the observed neutrino signal attributed to NGC~4151. In this work, we investigate the possible contributions of these nearby blazars to the neutrino excess close to NGC~4151 by numerical modeling of the observed multi-wavelength emission of 4FGL~J1210.3+3928 and 4FGL~J1211.6+3901. We discuss the detectability of the predicted neutrino flux from the two blazars by next-generation neutrino telescopes. Throughout this paper, we set the cosmological constants as $H_0 = 70$ km s$^{-1}$ Mpc$^{-1}$ and $\Omega_{\textrm{m}} = 0.3$.

\section{Data}

We collect the multi-wavelength data to build the spectral energy distribution of 4FGL~J1210.3+3928 and 4FGL J1211.6+3901.

\subsection*{4FGL~J1210.3+3928}
4FGL~J1210.3+3928 is a BL Lac object located at redshift $z=0.615$ \citep{1991ApJS...76..813S}. We obtain gamma-ray data from the Fermi-LAT 14-Year Point Source Catalog \citep[4FGL-DR4;][]{ballet2024fermi, Abdollahi_2022}. 4FGL~J1210.3+3928 is detected with $\sim 6 \sigma$ significance in gamma rays. The source has a gamma-ray flux of $(9 \pm 2) \times 10^{-11}~\mathrm{erg~s^{-1}}$ in the energy range of 1-100\,GeV. It is well described by a power-law spectral shape with a photon index of $2.1\pm0.2$. Its variability index is 14.42.

4FGL~J1210.3+3928 is a long-term monitored source in the X-ray band due to its proximity to NGC~4151 \citep{2008A&A...479...35M}. In X-rays, 4FGL~J1210.3+3928 was part of the field of view of 33 \textit{XMM-Newton} observations between 2000 and 2022. However, due to the chosen science mode of the observations, spectra could only be extracted for 11 observations. The science products are obtained using the \textit{XMM-Newton} Science Analysis System (SAS, Version 21.0.0). The spectra are later binned to have at least 20 counts in each bin. The extracted spectra show a flux variability of a factor 2 to 3 over the entire XMM observing period. For this study, we select a 10~ks observation on the 10th of December 2012 (obsID: 0679780401). This choice is motivated by two points: 1) selecting a high flux spectrum aligns with our goal to estimate the maximum neutrino emission of the source, 2) the observation date was also close to the rest of the multi-wavelength data gathered. The X-ray spectrum is well reproduced ($\chi^2/\textrm{dof}=26.5/22$) by a weakly absorbed power-law with a Hydrogen column density $N(H)=5.5^{+3.3}_{-3.2}\times10^{20}$ cm$^{-2}$ in agreement with the Galactic absorption level \citep{HI4PI2016} and a soft spectral index of $\Gamma=2.19^{+0.12}_{-0.12}$. The absorbed X-ray flux in the range 0.5 to 10 keV is $3.8\times10^{-12}~\mathrm{erg~cm^{-2}s^{-1}}$. The spectrum is then corrected for absorption in preparation for the modeling.

For the rest of the multi-wavelength data, optical data was taken from the SDSS Photometric Catalogue, Release 9 \citep{2012ApJS..203...21A}. Infrared emission was measured by the Wide-field Infrared Survey Explorer (WISE) and the corresponding fluxes were taken from the WISE All-Sky Data Release \citep{2012yCat.2311....0C, 2012wise.rept....1C}. The source was also detected in 1.4 GHz by the FIRST Survey in 1993 \citep{1997ApJ...475..479W}. 

\subsection*{4FGL~J1211.6+3901}
A second blazar, 4FGL~J1211.6+3901, is classified as BL Lac object as well (\citealt{Rector_2000}) with a redshift of $z=0.89$ based on spectroscopic measurements by \citet{10.1093/mnras/stu2665}. 4FGL J1211.6+3901 is detected in gamma rays with $\sim 5.4 \sigma$ significance \citep{ballet2024fermi, Abdollahi_2022}. The source has a gamma-ray flux of $(7 \pm 2) \times 10^{-11} ~\mathrm{erg~s^{-1}}$ in the energy range of 1-100 GeV and variability index of 15.52. The spectral shape follows a power law with a photon index of $1.8\pm0.2$.

 We analyze the data obtained during the 15ks  observation on the 14th of June 2002 by \textit{XMM-Newton}. The blazar, which was not the main target of the observation, was at the time in an enhanced state. \textit{XMM-Newton} data were obtained with the EPIC cameras \citep[][]{Struder2001,Turner2001} in extended full-frame window mode with the medium filter applied. Science products are obtained using the \textit{XMM-Newton} Science Analysis System (SAS, Version 21.0.0). The spectrum is later binned to have at least 20 counts in each bin. The background is found to dominate the spectrum above 7 keV, therefore we discard the energies above. The spectrum is first analyzed using \textsc{XSPEC} \citep[][]{Arnaud1996}. It is well reproduced ($\chi^2/\textrm{dof}=35.7/43$) by a weakly absorbed power-law with a Hydrogen column density $N(H)=7.9^{+4.2}_{-3.9}\times10^{20}$ cm$^{-2}$ in agreement with the Galactic absorption level \citep{HI4PI2016} and a soft spectral index $\Gamma=2.3^{+0.18}_{-0.17}$. The absorbed flux in the range 0.5 to 10 keV is $1.4\times10^{-12}~\mathrm{erg~cm^{-2}s^{-1}}$ . The spectrum is then corrected for absorption in preparation for the modeling. Unfortunately, the target falls outside the field of view of the Optical Monitor \citep[][]{Mason2001} on-board \textit{XMM-Newton}, and no simultaneous UV data point could be extracted.

Optical measurements were done by OSIRIS/R5000R \citep{10.1093/mnras/stu2665}, Catalina Sky Survey \citep{2009ApJ...696..870D}, and the SDSS Photometric Catalogue, Release 12 \citep{2015ApJS..219...12A}. Infrared emission was measured by the Wide-field Infrared Survey Explorer (WISE) and corresponding fluxes were taken from WISE All-Sky Data Release \citep{2012yCat.2311....0C, 2012wise.rept....1C}. The source was also detected in 1.4 GHz by the FIRST Survey in 1993 \citep{1997ApJ...475..479W}.

\section{Numerical Modeling}
\begin{figure}
   \centering
   \includegraphics[width=10cm]{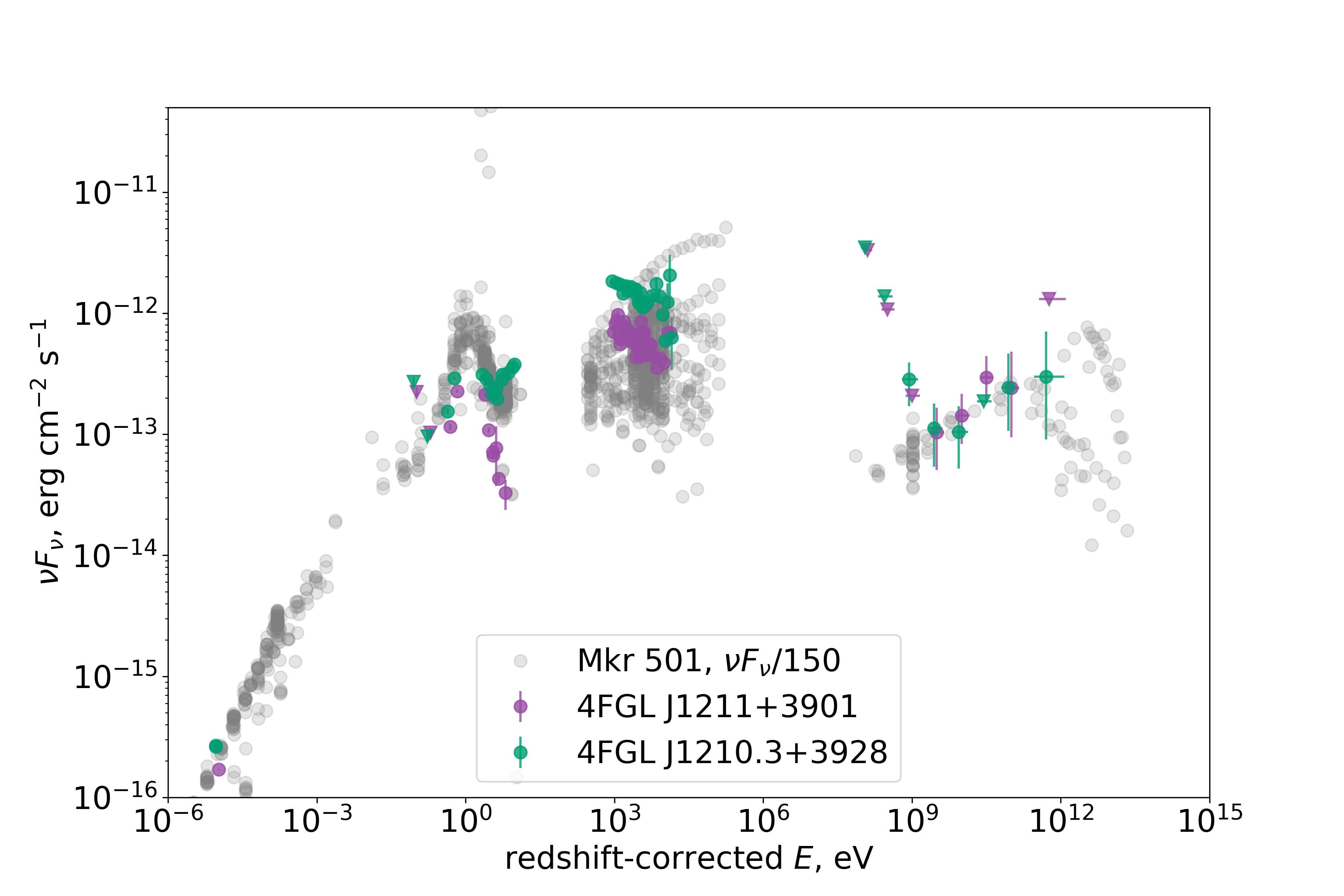}
      \caption{Comparison between spectral energy distributions of 4FGL~J1210.3+3928, 4FGL~J1211.6+3901 and Mrk~501. The grey data points correspond to the archival Mrk 501 SED scaled by 150. The green and purple data points (round -- measurements, triangles -- upper limits) correspond to 4FGL~J1211.6+3901 and 4FGL~J1211.6+3901 SEDs correspondingly.} 
         \label{fig:mrk501}
   \end{figure}

The spectral energy distribution (SED) of both 4FGL~J1210.3+3928 and 4FGL~J1211.6+3901 exhibit a black-body-like bump feature in the eV range, which does not show any significant variability \citep{2012yCat.2311....0C, 2012wise.rept....1C}. We consider that this feature is caused by the stellar emission from their host galaxies. A similar spectral feature is observed in the SED of Mkr 501, a prototype high-synchrotron-peaked BL Lac (HBL) object. A comparison of the rest frame SEDs of these blazars with Mrk 501\footnote{The Mrk 501 SED data was obtained through the SEDBuilder tool \url{https://tools.ssdc.asi.it/SED/}} is shown in Fig.~\ref{fig:mrk501}. The similarity in their broad-band spectral properties confirms the classification of both sources as typical HBLs. The eV bumps in both blazars are well reproduced by an elliptical galaxy spectral template with a stellar mass of $M=10^{12} M_\odot$. We adopt the stellar population synthesis model by \citep{2003MNRAS.344.1000B} assuming the Salpeter initial mass function, instantaneous star formation, age of 10~Gyr, and solar metallicity \citep[see][for details]{Itoh:2020qpn}.

\begin{figure} 
   \centering
   \includegraphics[width=10cm]{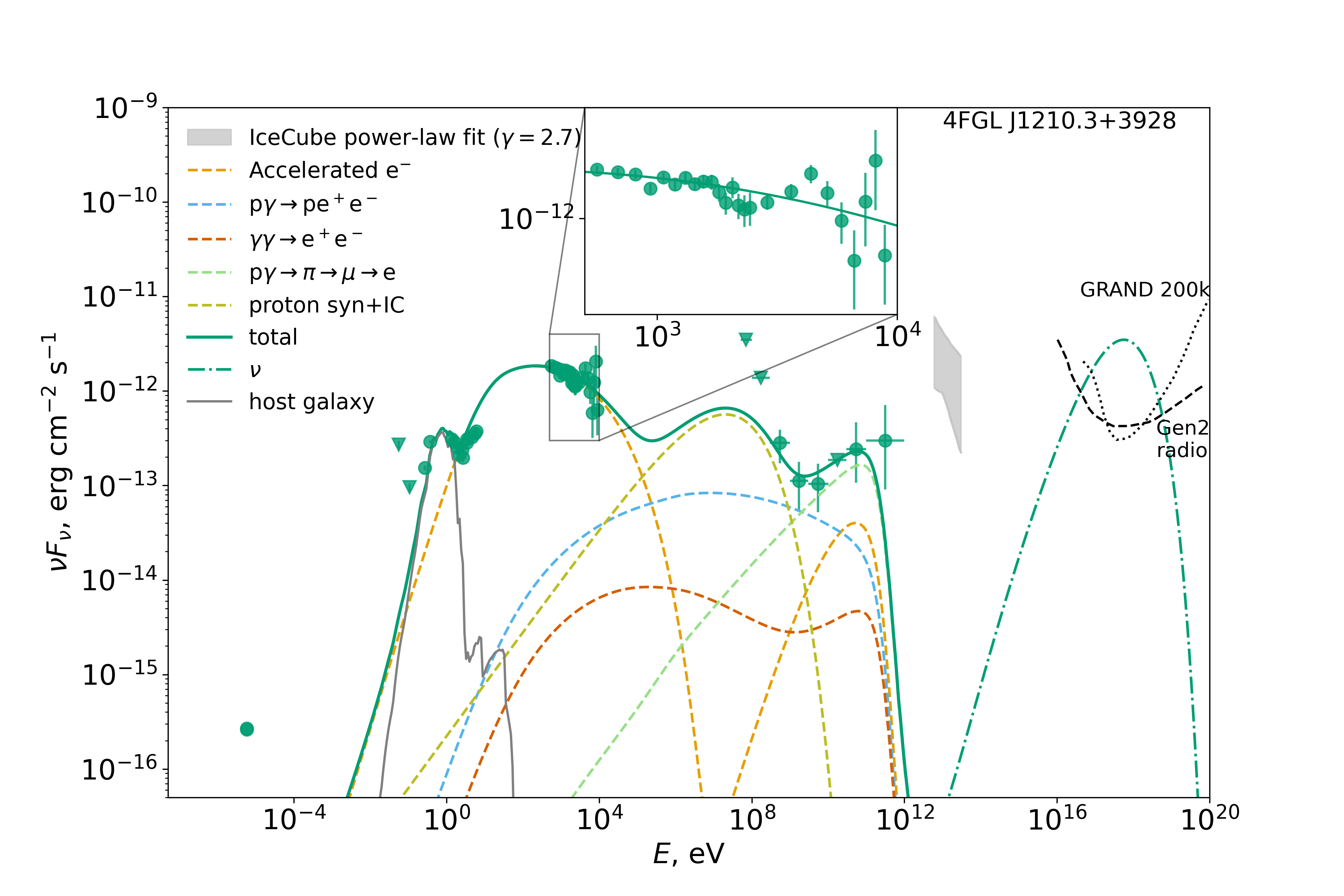}
      \caption{Leptohadronic model for 4FGL~J1210.3+3928. The green solid line corresponds to the multi-wavelength photon emission. The components are shown with the dashed line. The green dash-dotted line corresponds to the all-flavor neutrino spectrum. The gray shaded area is a neutrino flux from NGC 4151 under the assumption of power-law with spectral index $\gamma = 2.7$ \citep{2024arXiv240607601A}. The black dash-dotted line shows 10 year sensitivity of IceCube-Gen2 optical array \citep{2021JPhG...48f0501A}, black dashed -- 10 year all-flavor sensitivity of IceCube-Gen2 radio array \citep{2021JPhG...48f0501A}, black dotted line -- 10 year all-flavor sensitivity of GRAND 200k \citep{GRAND:2018iaj}.
              }
         \label{fig:lephad_model_J1210}
   \end{figure}

\begin{figure} 
   \centering
   \includegraphics[width=10cm]{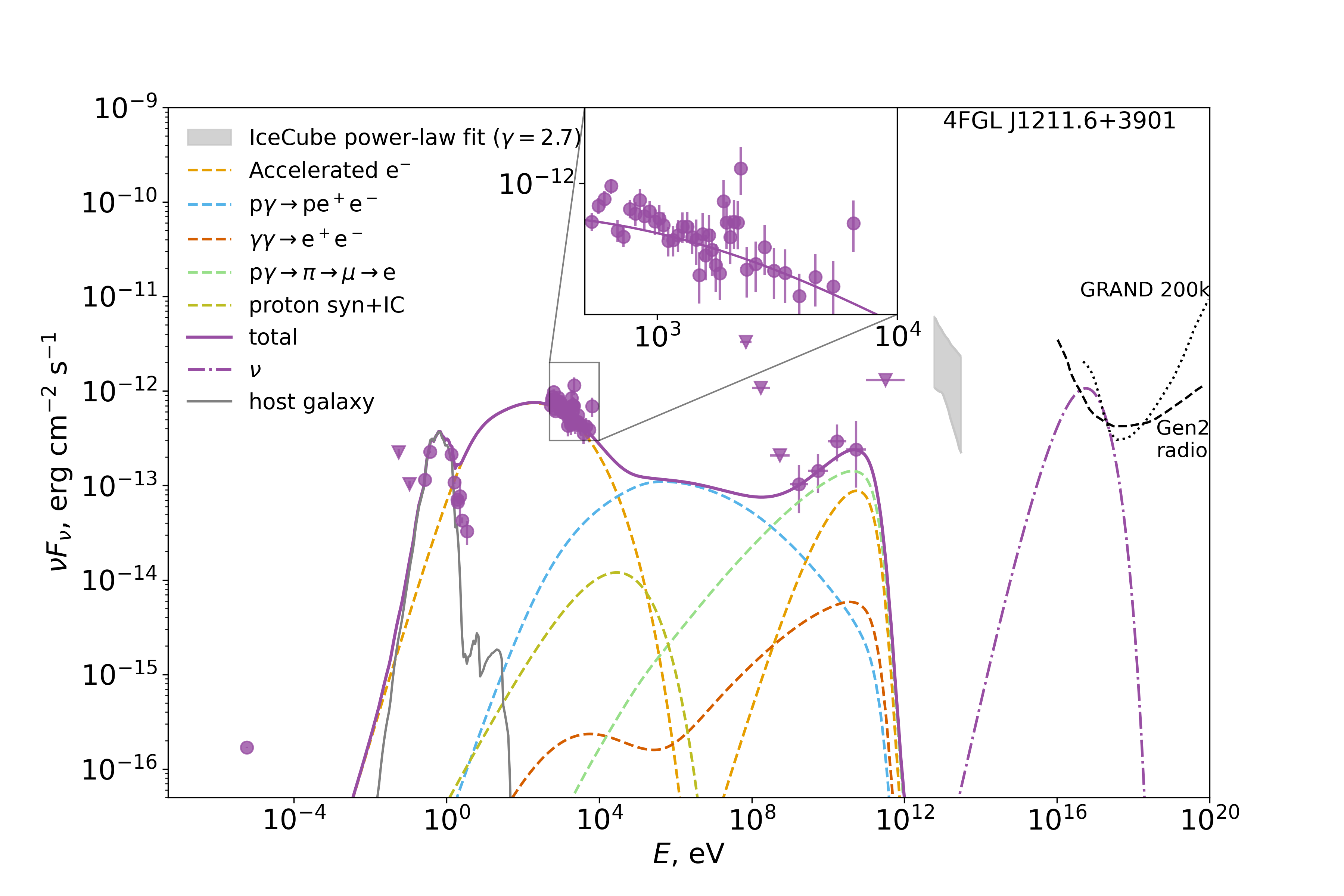}
      \caption{Leptohadronic model for 4FGL~J1211.6+3901. The purple solid line corresponds to the multi-wavelength photon emission. The components are shown with the dashed line. The purple dash-dotted line corresponds to the all-flavor neutrino spectrum. The gray shaded area is a neutrino flux from NGC 4151 under the assumption of power-law with spectral index $\gamma = 2.7$ \citep{2024arXiv240607601A}. The black dash-dotted line shows the 10 year sensitivity of the IceCube-Gen2 optical array \citep{2021JPhG...48f0501A}, black dashed -- 10 year all-flavor sensitivity of IceCube-Gen2 radio array \citep{2021JPhG...48f0501A}, black dotted line -- 10 year all-flavor sensitivity of GRAND 200k \citep{GRAND:2018iaj}.
              }
         \label{fig:lephad_model_J1211}
   \end{figure}

For each state, we numerically model multi-wavelength emission using the time-dependent code AM$^3$ \citep{klinger2023am3}, which solves a system of coupled differential equations describing the transport of particles interacting in the jet in a self-consistent way. Motivated by the possible neutrino emission, we start with a model where all radiation above $\sim$10 eV originates from radiation processes of electrons and protons in the jet. We assume that both electrons and protons are accelerated in the source to power-law spectra\footnote{Parameters with or without prime refer to the values in the jet or observer's frame respectively.}  $dN/d{\gamma'}_\mathrm{e,p} \propto {\gamma'}_\mathrm{e,p}^{-\alpha_\mathrm{e,p}}$ with spectral indices $\alpha_\mathrm{e,p}$, spanning a range of Lorentz factors from  ${\gamma'}_\mathrm{e,p}^\mathrm{min}$ to ${\gamma'}_\mathrm{e,p}^\mathrm{max}$. The energy spectra of the electrons and protons are normalized to the corresponding total electron and proton luminosities, $L'_{\mathrm{e}}$ and $L'_{\mathrm{p}}$.  These particles are then injected into a single spherical blob of size $R'$ (in the comoving frame of the jet) moving along the jet with Lorentz factor $\Gamma$, where there is a homogeneous and isotropic magnetic field of strength $B'$. We assume the jet is observed at an angle $\theta_{\textrm{obs}} = 1/\Gamma_{\textrm{b}}$ relative to its axis, resulting in a Doppler factor of $\delta_{\textrm{D}}\approx\Gamma_{\textrm{b}}$. We account for the high-energy gamma-ray absorption due to extragalactic background light (EBL) using the model by \cite{2011MNRAS.410.2556D}. The best-fit parameter values were found by minimizing the reduced $\chi^2$ with the Minuit package \citep{1975CoPhC..10..343J}. The results of the leptohadronic modeling for 4FGL~J1210.3+3928 and 4FGL~J1211.6+3901 are shown in Fig.~\ref{fig:lephad_model_J1210} and Fig.~\ref{fig:lephad_model_J1211} respectively. We show all-flavor neutrino fluxes here. The values of the leptohadronic parameters can be found in Table~\ref{table:pars}.

\begin{figure}
   \centering
   \includegraphics[width=10cm]{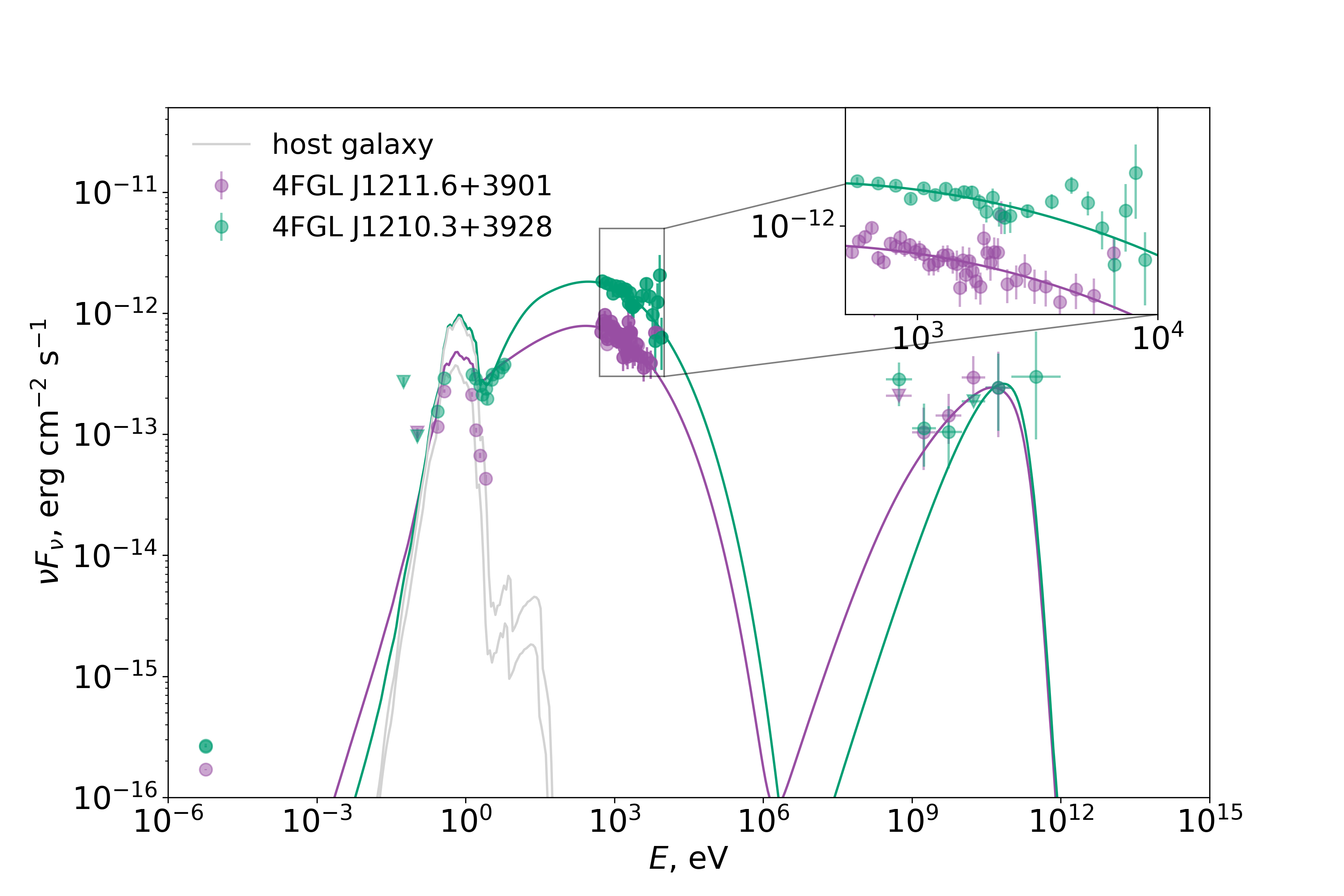}
      \caption{Leptonic models for both quiescent (blue) and flaring (red) states. The solid line corresponds to the multi-wavelength photon emission. The solid gray line corresponds to the emission of the host galaxy.
              }
         \label{fig:lep_models}
   \end{figure}

As an alternative scenario, we also consider a case where all radiation originates from purely leptonic processes. We assume that electrons are accelerated to a single power-law spectrum $dN/d{\gamma'}_\mathrm{e} \propto {\gamma'}_\mathrm{e}^{-\alpha_\mathrm{e}}$ with spectral index $\alpha_\mathrm{e}$, spanning a range of Lorentz factors from  ${\gamma'}_\mathrm{e}^\mathrm{min}$ to ${\gamma'}_\mathrm{e}^\mathrm{max}$. The energy spectrum of the electrons is normalized to the total electron luminosity parameter, $L'_{\mathrm{e}}$. Similarly to the leptohadronic case, electrons undergo interactions and radiate inside of a spherical blob of size $R'$ with a homogeneous and isotropic magnetic field of strength $B'$ moving along the jet with Lorentz factor $\Gamma$. The obtained best-fit solutions for both sources are shown in Fig. \ref{fig:lep_models}.

\section{Results}

The best-fit values of the leptohadronic models for both sources have typical values for HBL sources \citep{2024A&A...681A.119R}. The characteristic size of the emission zone is $5 \times 10^{16}$ cm with a magnetic field strength of 0.05 -- 0.1 G. Both sources require the presence of a high-energy electron population (with Lorentz factors ranging from ${\sim}10^4$ to ${\sim}10^6$) in order to explain the optical and X-ray fluxes. To produce neutrinos, highly relativistic protons must be present in the jet. The maximum proton energy in our models reaches ${\sim}10^{18}$ eV for 4FGL~J1210.3+3928, which corresponds to the ultra-high-energy cosmic ray (UHECR) regime, while it is ${\sim}10^{17}$ eV for 4FGL~J1211.6+3901. Similar values for the maximum proton energies were predicted for a different HBL, Mkr 421, in \cite{2014APh....54...61D}.

The gamma-ray spectrum of 4FGL~J1210.3+3928 exhibits a distinctive dip in the GeV range (Fig.~\ref{fig:lephad_model_J1210}). This spectral feature is successfully reproduced by our leptohadronic model, which combines proton-synchrotron emission ($\lesssim$ GeV) of with photopion production ($\gtrsim$ GeV). In contrast, purely leptonic scenarios fail to explain this spectral characteristic through inverse-Compton emission alone (Fig.~\ref{fig:lep_models}). 

In the case of 4FGL~J1211.6+3901, both leptonic and leptohadronic models describe the observed gamma-ray, X-ray and near-infrared fluxes well. The discrepancy in the optical band for both models can be caused by non-simultaneous data, which is important to account for effects of source variability. Unfortunately, with the limited optical data, no simultaneous SED could be built. We note that with the currently available data, there are no possibilities to discriminate between purely leptonic and leptohadronic scenario. An important energy region that can shed light on this problem is MeV gamma rays. High-energy photons passing in the vicinity of protons produce Bethe-Heitler pairs, which in turn also emit synchrotron radiation which peaks in the MeV range. This excess of MeV photons can be a hadronic signature which highlights the importance of future MeV Compton telescopes such as COSI \citep{2019BAAS...51g..98T}, GRAMS \citep{2020APh...114..107A}, and AMEGO-X \citep{2022JATIS...8d4003C}. 

In all of the presented models, the intrinsic gamma-ray luminosity, before EBL absorption, is comparable to the neutrino luminosity, maintaining energy conservation in the hadronic processes. However, in the observed spectra (as shown in Figs.~\ref{fig:lephad_model_J1210} and \ref{fig:lephad_model_J1211}), gamma rays with energies above TeV are strongly absorbed due to EBL, creating an apparent energetic discrepancy between gamma-ray and neutrino fluxes. We note that the electromagnetic cascade of these absorbed gamma rays could contribute to the observed MeV -- GeV fluxes. However, as the intergalactic magnetic field (IGMF) has a strength $> 7.1 \times 10^{-16}$ G \citep[e.g.,][]{2023ApJ...950L..16A}, the cascade emission can significantly dissipate. This is consistent with the lack of clear evidence for secondary cascade components in current gamma-ray observations. Given these observational constraints and the uncertainty of the IGMF strength, we did not include the intergalactic cascade component in our analysis. 

Our leptohadronic modeling predicts all-flavor neutrino fluxes of $\sim10^{-12}~\mathrm{erg~cm^{-2}~s^{-1}}$ peaking above $\sim$10~PeV for both blazars. These predicted fluxes suggest that the contribution of these sources to the neutrino signal observed by IceCube in the direction of NGC~4151 is minor.  In our data analysis, we selected a high level of X-ray flux to estimate the maximum possible neutrino contribution from both blazars. The time-averaged emission state would only further reduce blazar contribution to the NGC 4151 hotspot, strengthening our conclusion about its subdominant role.

Despite the improved PeV-range sensitivity of the optical array in the next-generation neutrino observatory IceCube-Gen2 \citep{2021JPhG...48f0501A}, the detection of possible neutrinos emission from 4FGL~J1210.3+3928 or 4FGL~J1211.6+3901 remains challenging due to their high energies. Detection of neutrinos at $10^{16} -10^{17}$~eV, where the emission peaks, requires either the radio array of IceCube-Gen2 or a detector like GRAND \citep{GRAND:2018iaj}. The predicted neutrino emission from 4FGL~J1210.3+3928 could be detectable by both facilities. The neutrino flux from 4FGL~J1211.6+3901 lies at the edge of GRAND~200k's 10-year sensitivity and slightly above the 10-year sensitivity of IceCube-Gen2's radio array, suggesting possible detection with IceCube-Gen2. We note that the accuracy of the directional reconstruction is expected to be better than 1$^\circ$ for the IceCube-Gen2 radio array \citep{2021JPhG...48f0501A} and better than 0.5$^\circ$ for GRAND \citep{GRAND:2018iaj}. In case of the detection of multiple neutrinos, these values can be further improved, leading to a possible spatial separation of signals from 4FGL~J1210.3+3928 and 4FGL~J1211.6+3901.

\section{Discussion and Conclusions}

The IceCube analysis has found around 30 signal neutrinos from NGC 4151 in 10 years of data \citep{2024arXiv240607601A}. However, two gamma-ray bright blazars, 4FGL~J1210.3+3928 and 4FGL~J1211.6+3901, are located 0.08$^\circ$ and 0.43$^\circ$ from NGC 4151 respectively. Blazars are promising neutrino emitters and can contribute to the observed IceCube signal. We modeled the multi-wavelength spectrum of both blazars. The leptohadronic model of 4FGL~J1210.3+3928 explained the observed electromagnetic fluxes including a GeV dip in gamma rays, which can be explained by a purely leptonic model. For 4FGL~J1211.6+3901, both leptonic and leptohadronic models explain the observed data equally well. We found that the predicted neutrino flux peaks around $10^{17}$ eV for both sources. The contribution to the observed TeV neutrino flux is expected to be subdominant. 

Both 4FGL~J1210.3+3928 and 4FGL~J1211.6+3901 have high-energy synchrotron peaks in X-ray, indicating the presence of efficient particle acceleration. In our model, the neutrino spectrum peaks at $\sim$10--100~PeV due to $p\gamma$ interactions with internal synchrotron photons and the extension of cosmic-ray spectra to EeV energies. While lowering the maximum proton energy could shift the neutrino peak toward TeV energies \citep{2014JHEAp...3...29D}, where IceCube detected neutrinos from the direction of NGC~4151, such models would require substantially higher proton powers and fail to explain the observed broad-band electromagnetic spectra. The maximum proton Lorentz factor for the leptohadronic model of 4FGL~J1211.6+3901 is substantially lower than that for 4FGL~J1210.3+3928. This is caused by the fact that with the increase of proton energy, the contributions of Bethe-Heilter pairs and proton synchrotron become more significant leading to the violation of the observed  X-ray spectrum and featureless gamma-ray spectrum.This demonstrates the robustness of our solution within the large parameter space of one-zone radiation models, despite potential degeneracies in the model parameters \citep{2024arXiv240904165O}.

The proton luminosity is fundamentally limited by the accretion power. Since we do not have the measurements of accretion powers in these blazars, we follow the Eddington power argument. We can do an order-of-magnitude Eddington luminosity estimation using the relation between black hole mass and bulk mass (e.g. \citealt{2004ApJ...604L..89H, Zhu_2021}). Based on our fitted galactic profile, $M_{\textrm{bulk}} = 10^{12} M_\odot$ which roughly corresponds to $M_{\textrm{BH}} = 10^9 M_\odot$ leading to $L_{\textrm{Edd}} \sim 10^{47}$ erg/s. When lowering the maximum proton energy to $10^{15} - 10^{16}$ eV and setting the proton luminosity to the Eddington luminosity, the neutrino spectrum is still expected to peak above PeV energies and thus the blazar contribution to the neutrino signal from the region near NGC 4151 can be only subdominant.

Models for neutrino emission from the Seyfert galaxy NGC~4151 predict neutrinos at lower energies compared to our models for the two nearby blazars. While most of the models predict a cut-off of the neutrino spectrum above 10-100 TeV for Seyfert galaxies, HBLs can produce neutrinos at higher energies. Next-generation neutrino telescopes such as IceCube-Gen2 or GRAND may solve this discrepancy by probing a higher energy range and possibly spatially discriminating two hotspots. In addition, future MeV gamma-ray missions will be able to test signatures of hadronic emission.

\begin{acknowledgements}

      The authors thank Anna Franckowiak for useful comments and discussions. AO is supported by DAAD funding program 57552340 and RUB Research School. AO acknowledges the support from the DFG via the Collaborative Research Center SFB1491 Cosmic Interacting Matters - From Source to Signal. SB is an overseas researcher under the Postdoctoral Fellowship of Japan Society for the Promotion of Science (JSPS), supported by JSPS KAKENHI Grant Number JP23F23773. YI is supported by NAOJ ALMA Scientific Research Grant Number 2021-17A; by World Premier International Research Center Initiative (WPI), MEXT; and by JSPS KAKENHI Grant Number JP18H05458, JP19K14772, and JP22K18277.
\end{acknowledgements}

%
%

\bibliography{J1211}

\clearpage

\begin{appendix}
\section{Model parameters}

\begin{table}[H]
\caption{Table of best-fit parameters for leptonic and leptohadronic models during the quiescent and flaring states of J1211.6+3901}

\begin{tabular}{lcccc}
\toprule
Parameters     & \makecell{4FGL J1210.3+3928 \\leptohadronic}   & \makecell{4FGL J1211.6+3901 \\leptohadronic}  & \makecell{4FGL J1210.3+3928 \\leptonic} & \makecell{4FGL J1211.6+3901\\ leptonic}   \\
\midrule
$R'_{\mathrm{b}}$ [cm]      & $5 \times 10^{16}$ & $5 \times 10^{16}$ & $3 \times 10^{16}$   & $3.5 \times 10^{16}$ \\
$B'$ [G] & 0.1 & 0.04 & 0.05   & 0.04\\

$\Gamma_{\mathrm{b}}$   & 25 & 25 & 25   & 25  \\
$\gamma'_{\mathrm{e, min}}$   & $2 \times 10^4$ & $2 \times 10^4$ & $2.5 \times 10^4$   & $1 \times 10^4$ \\
$\gamma'_{\mathrm{e, max}}$   & $1 \times 10^6$ & $1 \times 10^6$ & $1 \times 10^6$   & $1 \times 10^6$ \\

$\alpha'_{\textrm{e}}$   & 2.0 & 2.3 & 2.3   &2.3 \\

$L'_{\mathrm{e}}$ [erg/s]& $3 \times 10^{40}$ & $7.5 \times 10^{40}$ &  $7\times 10^{40}$&  $1.4 \times 10^{41}$\\

$\gamma'_{\mathrm{p, min}}$   & $2 \times 10^3$ &$2 \times 10^3$ & --  &  -- \\
$\gamma'_{\mathrm{p, max}}$   & $1.5 \times 10^9$ &$6 \times 10^7$ & --  & -- \\

$\alpha'_{\mathrm{p}}$   & 1.9 & 1.5 & --   & -- \\

$L'_{\mathrm{p}}$ [erg/s]& $4 \times 10^{45}$ & $1.5 \times 10^{46}$ &  -- & -- \\

\bottomrule

\end{tabular}

\label{table:pars}
\end{table}

\end{appendix}

\end{document}